\documentclass{ws-procs9x6}

\begin{document}

\begin{center}
To appear on the Proceedings of the 14th ICATPP Conference on\\
Astroparticle, Particle, Space Physics and Detectors\\ for Physics Applications,\\ Villa  Olmo (Como, Italy), 23--27 October, 2013, \\to be published by World Scientific (Singapore).
\end{center}
\vspace{-1.7cm}

\title{Possible Contribution to Electron and Positron Fluxes from Pulsars and their Nebulae}
\author{S. Della Torre$^{1}$, M. Gervasi$^{1,2}$, P.G. Rancoita$^{1}$, D. Rozza$^{1,3,4,*}$, A. Treves$^{1,3}$.}
\address{
$^1$ \textit{INFN Sezione di Milano Bicocca, I-20126 Milano, Italy}\\
$^2$ \textit{Universit\`a di Milano Bicocca, I-20126 Milano, Italy} \\
$^3$ \textit{Universit\`a dell'Insubria, I-22100 Como, Italy}\\
$^4$ \textit{European Organization for Nuclear Research, CERN, CH-1211 Geneva 23, Switzerland}\\
$^*$ Corresponding Author: davide.rozza@mib.infn.it
}
\begin{abstract}
The AMS-02 experiment confirms the excess of positrons in cosmic rays (CRs) for energy above 10 GeV with respect to the secondary production of positrons in the interstellar medium. This is interpreted as evidence of the existence of a primary source of these particles. Possible candidates are dark matter or astrophysical sources. In this work we discuss the possible contribution due to pulsars and their nebulae. Our key assumption is that the primary spectrum of electrons and positrons at the source is the same of the well known photon spectrum observed from gamma-rays telescopes. Using a diffusion model in the Galaxy we propagate the source spectra up to the Solar System. We compare our results with the recent experiments and with the LIS model.
\end{abstract}
\bodymatter

\section{Introduction}
Electrons and positrons represent $\sim1\%$ of the total amount of cosmic rays at $\sim10$ GeV. Electrons are produced by primary sources like supernova remnants and by secondary production from interactions between cosmic protons and light nuclei with the interstellar medium (ISM). Cosmic positrons are supposed to be produced only in secondary production. Under these informations the positron fraction (ratio between the positron and the sum of the positron and electron fluxes), for energy above 10 GeV, should decrease when the energy increase \cite{Delahaye2010}\!. The recent result published by the AMS-02 collaboration confirms the positron excess in cosmic rays \cite{AMS02posfrac}\!. The model that predicts only the secondary production for positrons does not account the experimental data. The positron fraction may be explained by primary sources e.g. dark matter or pulsars and their nebulae. We focused our attention on the second possibility.

\section{Local Interstellar Spectrum of Electrons and Positrons in Cosmic Ray}
The propagation of positrons and electrons in the interstellar medium may be described by the following diffusion equation\cite{Malyshev09}\!.
\begin{eqnarray}\label{EqDiff}
  \frac{\partial N_{e}(\vec{x},E,t)}{\partial t} & = & Q(E)\nonumber\\
  & & +\vec{\nabla}\cdot\left[D(E)\vec{\nabla}N_{e}(\vec{x},E,t)\right]\\
  & & +\frac{\partial}{\partial E}\left[b(E)N_{e}(\vec{x},E,t)\right]\nonumber
\end{eqnarray}
The energy density $N_{e}$ of the particle as a function of the time depends on sources $Q(\vec{x},E)$, diffusion ($D(E)$ is the diffusion coefficient) and energy loss terms. The last one is very important for electron like particles that interact with the interstellar medium and with the magnetic and radiation fields \cite{Schlickeiser}\!. Ionization energy loss increases with the logarithm of the energy, the Bremsstrahlung term is linear while synchrotron and inverse Compton terms increase with the square of the energy. Above 1 GeV the main contribution to energy loss is due to these last two terms\cite{Yoshida} $b(E)\sim b_0E^2$.
\newline
Using a diffusion coefficient with a power law of the energy and a break at $E_1$:
\begin{equation}
 D(E)=D_{0}\left(\frac{E}{E_{1}}\right)^{\delta}
\end{equation}
it is possible to estimate the distance ($R\sim\sqrt{2D(E)\tau}$) reached by a particle after a time $\tau$. If we consider a positron of 1 TeV, the time for lose all the energy is about $10^5$ years and the distance is $\sim1$ kpc. In this case if we want to search a source of this particles at high energy we need to focus our attention in a region relatively close to us, less than 2 kpc.\\
To estimate the primary contribution in the electron and positron fluxes we need to know the local interstellar spectrum (LIS) for the different species of particles using the GALPROP code \cite{galpropweb,vladimirov}\!. We optimized the diffusion parameters to reproduce fluxes and ratios in cosmic rays. The energy dependence of the propagation in the ISM is commonly investigated with measurements of the abundance of secondary cosmic-ray nuclei produced in spallation processes (like boron), relative to the abundance of their parent primary cosmic-ray species (like carbon). These studies provide a direct measurement of the Galactic propagation path length of cosmic rays and consequently to the diffusion coefficient. It has been found that the path length decreases with energy, perhaps in form of a power law \cite{Malyshev09}\!. Analysing the data of the ratio B/C and other secondary over primary CRs in the experiments from 1990 up to now, we found an agreement using, for GALPROP, a diffusion constant of $D_{0}=5.8\cdot10^{28}$ cm$^2$/s and $\delta=0.33$. Starting from the electron and proton fluxes presented this year from the AMS-02 collaboration\cite{icrc2013,icatppprocee} we chose an injection spectral index of 1.98, 2.42 below and above 9 GeV for nuclei and 1.70, 2.68 below and above 4 GeV for electrons. The GALPROP positron flux is computed from the interactions of the primary spectrum with the ISM. Fig.\ref{fig:signal} shows the difference between the electron and positron differential intensity observed by AMS-02 and the local interstellar spectrum for these particles from GALPROP. The error bars do not consider the band of variability of GALPROP parameters. The fluxes for both particles are comparable suggesting that they are created by sources of electron-positron pairs, maybe in electromagnetic shower. Moreover it is possible to observe the common power law, function of energy, that we are searching as primary contribution.

\section{Possible Primary Contribution from Pulsars and their Nebulae and Comparison with Experimental data}
\begin{figure}[t]
 \psfig{file=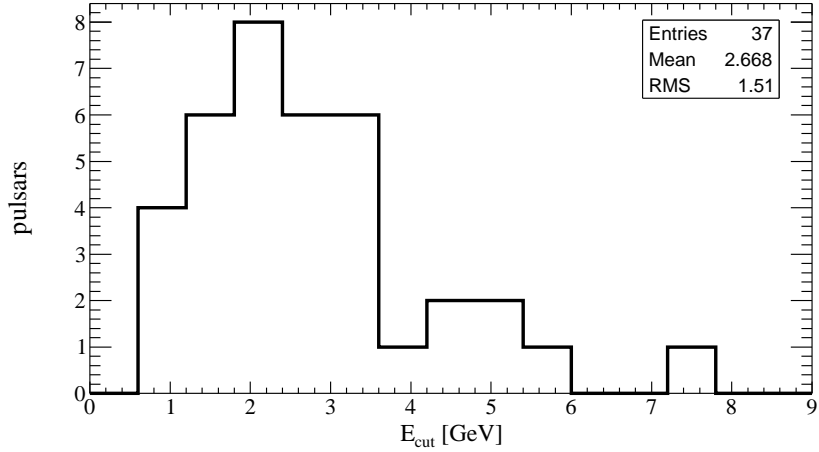,width=1\textwidth}
 \caption{Distribution of the energy cutoff for pulsars in the first Fermi catalogue.}
 \label{fig:cutoff}
\end{figure}

\begin{figure}[t]
 \psfig{file=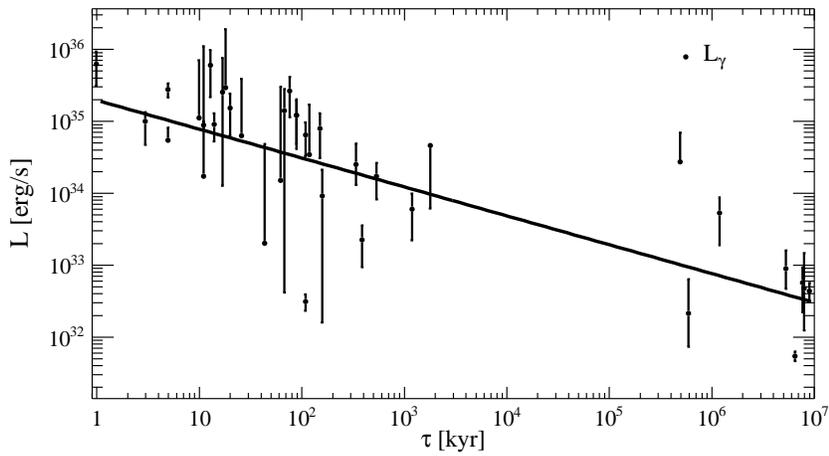,width=1\textwidth}
 \caption{Gamma luminosity vs. age for pulsars detected by Fermi-LAT \cite{Abdo2010}\!; in black line the fit.}
 \label{fig:lumi}
\end{figure}

\begin{figure}[t]
 \psfig{file=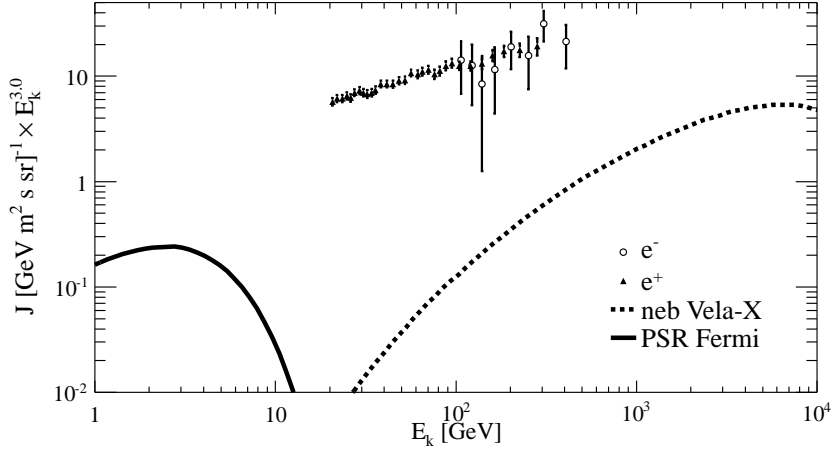,width=1\textwidth}
 \caption{The experimental data are the difference between the AMS-02 flux and the LIS for electrons (empty circle) and positrons (full triangle). In solid line the sum of the contributions of the 37 pulsars and in dashed line the contribution of Vela-X nebula.}
 \label{fig:signal}
\end{figure}

As so far implemented, pulsars and their nebulae may emit electrons and positrons in different regions. Our main assumption is that we used the well know observed experimental data of gamma spectrum of pulsars and nebulae as the primary spectrum for electrons and positrons at the source \cite{RozzaICRC}\!. Rescaling the spectrum of the gamma rays to the source we used the same shape (spectral index and energy cutoff) and the same luminosity for the electrons and positrons spectrum. Our reference is the observed spectra of the 37 pulsars reported in the first Fermi LAT catalogue \cite{Abdo2010}\!. As reported by the Fermi collaboration, the pulsar spectra were fitted with a power law (spectral index $\alpha$) and an exponential energy cutoff $E_{cut}$ of the form:
\begin{equation}
 \frac{dN_{\gamma}(E)}{dE}=K E^{-\alpha}e^{\frac{E}{E_{cut}}}.
\end{equation}
$K$ is the normalization factor coming from the eq. (5-6) and table 4 of Ref. [11]. Fermi data report a peak in the spectral index distribution in the range between 1 and 2, but more important is that the cutoff is always less than 10 GeV and peaks around 1-3 GeV (we used the values reported in tables 1 and 4 of Ref. [11]). Fig. \ref{fig:cutoff} reports the distribution of the energy cutoff for these pulsars.
Comparing the gamma luminosity of the pulsars and their ages in a scatter plot we found a possible power law correlation between the age ($\tau$) and the gamma luminosity $L_{\gamma}(\tau)\propto \tau^{-b}$ with $b\sim0.4$ (Fig. \ref{fig:lumi}).
Having the luminosity as a function of the time it is possible to integrate it to estimate the total energy output for the normalization constant of the primary electron/positron spectrum \cite{Malyshev09,RozzaICRC}\!.
\begin{equation}\label{Qinjection}
 Q(E)=Q_{0}E^{-\alpha}e^{\frac{E}{E_{cut}}}\qquad ;\qquad Q_{0}=\frac{\int{L_{\gamma}(t)dt}}{\int{E^{-\alpha}e^{\frac{E}{E_{cut}}}dE}}
\end{equation}
In this calculation we assumed a geometrical correction factor equal to one for photons. This parameter is model dependent and depends on the inclination of the magnetic axis respect to the rotational one and the observer angle. In the common models that describe the particle emission from pulsars, this factor is about one (as reported in Ref. [11]). We maximised the one for particles using the same value of the gamma rays.\\
The AMS-02 experimental data reported a positron excess up to 350 GeV, knowing that the energy cutoff for pulsars is less than 10 GeV, we investigated also other possible contributions from the nebulae associated to the pulsars that can reach the TeV energy region like Vela-X nebula \cite{HESS06}\!. The informations about these sources are very poor. The pulsar wind nebula (PWN) sources emitting in the TeV region are 34 (e.g., see Ref. [14]). Most of them are located at a distance larger than 2 kpc (see discussion in Sect. 2) from the Solar System (and the particles can not reach the Earth position due to the energy loss) or it is unknown. Five sources were observed at a distance lower than 2 kpc. The PWN identified in the Geminga region is classified with too low significance\cite{Milagro2009} and, for this reason, is not considered in the present analysis. We do not have information about the spectrum of the PWN associated to the pulsar Boomerang. The other three sources, Vela-X, CTA1 and Crab, have the characteristics needed for the current study. Crab is 1000 years old, we can use this primary spectrum as the spectrum of the other two nebulae in the first thousand years of their life. We analysed the nebulae under the same assumptions made before for the pulsars and we propagated their spectra \cite{HESS06,HessCrab,FermiCrab,FermiVelaNeb,VeriCTA2013}\!. We found a correlation between the gamma luminosity of the three nebulae and their ages with an index $b\sim2$.\\
During the life of the sources we divided the spectrum in step of 1000 years. Each of these steps corresponds to a bunch of energy emitted at the end of the step. In this case the time between the end of the bunch and the present epoch is the diffusion time $t$. The integral over time in equation (\ref{Qinjection}) corresponds to the duration of the step (1000 years). The diffuse spectra for pulsars and nebulae are calculated following the analytic solution of the diffusion equation reported in Ref. [3,19]. We defined the flux $J(E)=cN_{e}(E)/(4\pi)$ and:
\begin{equation}\label{lambda}
\lambda_{d}^{2}(E,E_{0})=\int^{E_{0}}_{E}\frac{D(E')dE'}{b(E')}
\end{equation}
where $\lambda_{d}$ is the mean distance travelled by particles from energy $E_0$ down to energy $E$ considering the energy loss and the diffusion. The differential intensity for positrons or electrons injected from a source distant $\vec{x}$ and diffuse in the ISM for a time $t$ is:
\begin{equation}\label{flux}
J(\vec{x},E,t) = \frac{c}{4\pi}\frac{Q_0}{(4\pi\lambda_d^2)^{3/2}}E^{-\alpha}\left(1-b_0tE\right)^{\alpha-2}e^{-\frac{E}{E_{cut}(1-b_0tE)}}e^{-\frac{\vec{x}^{2}}{4\lambda_d^2}}
\end{equation}
Fig. \ref{fig:signal} shows the difference between the AMS-02 data and the LIS compared with the possible primary contribution of the 37 pulsars and the Vela-X nebula. At 300 GeV there are more than one order of magnitude between the data and the possible nebula flux.
The pulsar contribution is mainly due to Geminga that is very close to us ($\sim200$ pc); the total spectrum from pulsars is at low energy because all the pulsars have an energy cutoff less than 10 GeV and they can not explain the positron excess at $\sim100$ GeV and more. For this reason we introduced in this analysis also the spectra from the nebulae of the pulsars. Vela-X spectrum is the only that we can keep in account, CTA1 contribution is negligible (the nebula is distant 1.4 kpc) and the Crab spectrum is not reported because the time between the emission of the particles and the present time is less than the diffusion time. Most of these nebulae were observed by HESS telescopes in Namibia. The sky coverage of this experiment is about half of the Fermi one\footnote{Pulsars and nebulae are distributed in the Galactic plane and HESS can cover about half of this region.}. Thus, to a first approximation, the non-observed nebulae - due to the lack of sky coverage by HESS - may contribute to increase the current estimate to a factor 2 or 3 at most\footnote{The Boomerang nebula is in the region not observed by HESS and it is keep in account in this factor.}. Also in this cases the total flux from the nebulae is lower than the data. Moreover it is hard to explain the possible existence of other nebulae not related to the pulsar objects. In conclusion, if we want to interpreter the positron excess with pulsars we need to have for them a nebula that can emits at high energy (above tens GeV), but they are not actually observed.

\section{Conclusion}
In this work we analysed the LIS for electrons and positrons in cosmic rays to search for a primary contribution of these particles that does not coming from the interaction between CRs and the interstellar medium. We studied the possible contribution of primary particles from pulsars and their nebulae for interpreting the results of AMS-02. Under the assumption that the injection spectrum of electrons and positrons is the same observed in gamma rays, we compared the diffuse differential intensity with the experimental data. This contribution is hardly able to explain the observations of AMS-02 data.\\

\end{document}